# TEMPORAL CHARACTERIZATION OF THE REMOTE SENSORS RESPONSE TO RADIATION DAMAGE IN L2

*Ruben De March[1], Deborah Busonero[2], Rosario Messineo[1], Alessandro Bemporad[2], Francesco Vaccarino[3], Angelo Fabio Mulone[1], Andrea Fonti[1], Mario Lattanzi[2]*

[1]ALTEC S.p.A., Corso Marche 79, 10146 Torino, Italy
[2]INAF-OATo, Via Osservatorio 20, 10025 Pino Torinese, Torino, Italy
[3]Politecnico di Torino, Dept. Math. Sciences, Corso Duca degli Abruzzi 24, 10129 Torino, Italy

**ABSTRACT**

Remote sensors on spacecrafts acquire huge volumes of data that can be processed for other purposes in addition to those they were designed for. The project TECSEL2 was born for the usage of the Gaia AIM/AVU daily pipeline output and solar events data to characterize the response of detectors subjected to strong radiation damage within an environment not protected by the terrestrial magnetic field, the Lagrangian point L2, where Gaia operates. The project also aims at identifying anomalies in the scientific output parameters and relate them to detectors malfunctioning due to radiation damage issues correlating with solar events occurred in the same time range. TECSEL2 actually designs and implements a system based on big data technologies which are the state of art in the fields of data processing and data storage. The final goal of TECSEL2 is not only related to the Gaia project, because it provides useful analysis techniques for generic and potentially huge time series datasets.

## 1. INTRODUCTION

Gaia is the ESA space mission launched on December 19[th], 2013 from the French Guyana, aiming at Global Astrometry at few $\mu$as, scanning continuously the whole sky in order to build the largest, most precise three-dimensional map of our Galaxy by surveying more than a thousand million stars [1], [2]. It is the first mission to operate within the Charge Couple Devices bandwidth in L2. Gaia focal plane array with its 106 CCDs is therefore an invaluable source of information about the CCDs behaviour within a strong radiation environment.

The Astrometric Instrument Model (AIM) is one of the crucial components of the Astrometric Verification Unit (AVU), the verification counterpart operating independently from the Gaia baseline data reduction chain. The AIM system is devoted to the monitoring, diagnostic and calibration of the Gaia astrometric instrument response over the mission lifetime [3], [4]. AVU has its own dedicated Data Processing Centre in Turin, the DPCT, which is one of the six Gaia DPCs spread across Europe. DPCT is also designed to provide computation, storage, data access and operations services to Italian Gaia Science Community [5].

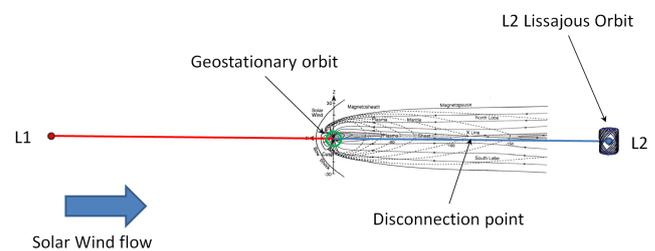

**Fig. 1**. L2 location vs. Earth's magnetosphere.

The availability of new big data technologies opens new scenarios in which science data collected inside a specific mission can be used to find new information and correlate it with datasets coming from different sources. The goal of TECSEL2 project is to study the possibility to use data acquired by remote sensors like CCDs to describe the environment in which the CCDs operate. The project starts with the preparation of data coming from Gaia CCDs in order to compare them with events related to the Sun.

## 2. RADIATION DAMAGE AND SCIENCE DATA

Solar flares and/or Coronal Mass Ejections (CMEs), quite often associated with the acceleration of Solar Energetic Particles (SEPs) released by the Sun [6], are the major responsible for disturbances on Earths magnetosphere and lead to Geomagnetic Storms. The explanations for these physical processes are far from being fully understood and are the main topic of the recently new-born Space Weather discipline.

In particular, one of the less known regions of Earth's magnetosphere is the magnetotail. The reason is that only a few satellites visited this far region and sampled the evolution of plasma parameters before, during and after Geomagnetic Storms, so further studies of the interplanetary environment in the far magnetotail are needed.

We need to keep in mind that the main contribution to the radiation damage is due to SEPs. During the maximum solar activity period of the solar cycle, solar eruptions produce large fluxes of solar energetic particles, protons and heavy ions, which reach L2 when the eruptions are directed towards Earth. Heavy ions will not contribute significantly, but may leave traces as strong transient events. The main contributor to transient events (detected cosmic ray traces) and accumulated radiation damage over the mission lifetime will be solar protons. The expectation is that transient effects and radiation damage increase can be correlated to solar flare events.

Further information on the radiation damage effects and evolution, and its impact on the read-out images will come from nominal processing of star images themselves during the mission lifetime. Science data can be used to trace directly the instrument response, taking advantage of the repeated measurements of stars over the CCDs field, and in particular to characterize the CTI effects.

Indeed, the radiation damage changes the population traps and the charge release, degrading the charge transfer efficiency. The consequent deformation of the PSF/LSF profile introduces a centroid bias and a potential flux loss, if not calibrated. Variation in time of critical parameters like centroid, background, flux and secondary processing outputs like diagnostic image profile parameters can be related to radiation damage effects.

## 3. DATA PROCESSING

### 3.1. Datasets

The input Gaia dataset consists of observations of stars that are processed by AIM module. Through the analysis we can subset by the CCD row data are coming from, the magnitude and/or the star object type. These data are thereafter processed (averaging values of different measures inside specific time intervals) to obtain time series from the several instant-provided values. The considered datasets are derived from observations of either stars brighter than magnitude 13 (for which bi-dimensional images are acquired) or stars whose magnitude ranges between 15 and 16, i.e. stars neither too bright nor too weak, in order to maximize the detection of the effect of radiation damage. Another choice is taking data from the first row of Gaia CCDs to have less disturbance of other types, like CCDs' background level ([7]).

The space environment measurements are provided by GOES-13 and GOES-15 satellites (placed in geostationary orbit) and by ACE and WIND spacecrafts (orbiting around L1 point). These data consist of time series of particles fluxes, subdivided by energy level and type of particle, and other measures, flow pressure for example. Moreover, the GOES-15 satellite provides X-ray flux data, useful to detect the occurrence of solar flares. These data are sampled by 1-minute, 5-minute or 1-hour intervals depending on the type of the series and on the satellite.

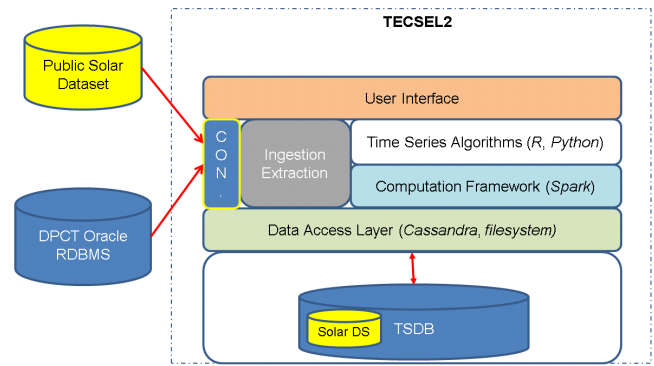

**Fig. 2**. TECSEL2 High Level Architecture

By the way, in the next Sec. 4 we will use the OMNI dataset, a dataset built by NASA with data from the spacecrafts mentioned above.

### 3.2. Architecture

TECSEL2 architecture is designed considering the described algorithms but also possible future integrations and other implementations. A high level description of the architecture can be seen in Fig. 2.

TECSEL2 system storage is populated through dedicated extraction/ingestion processes with solar datasets stored in files or with Oracle RDBMS used at DPCT for Gaia's data access and repository.

Since the algorithms belong to time series area, the system has to foresee a data storage component that uses a structure suitable for time series and provides scalable data access services independent from the number of time series and the data volume per series. In big data area this kind of storage system is called "time series database" (TSDB). The TSDB has been constructed comparing the available solutions and taking benefit from studies already conducted on the space sector [8].

TECSEL2 processing cluster currently consists of three nodes. This architecture takes the benefit of using Spark and Cassandra together, in particular performance is maximized when Spark runs on node where Cassandra has stored data Spark needs for its fast in-memory computations.

Given the sampling and the temporal range of the datasets, the total amount of objects stored inside Cassandra database and used in our analysis is about 20 million items.

### 3.3. Algorithms

Main computations involve cross-correlation and partial correlation between two paired samples, aiming at finding out if anomalous peaks (i.e., values evidently higher than usual ones) of solar wind cause any effect on Gaia CCDs. Moreover an incrementing time-shift is applied iteratively to one of the series to deduce, e.g., the delay between solar phenomena and their effect on Gaia. This delay depends on the time particles need to reach L2 point, but it can also depend on some additional time it takes for radiation damage to manifest itself. These algorithms pair datasets by entries' times, so they have to rightfully manage the instant times which lack of measures.

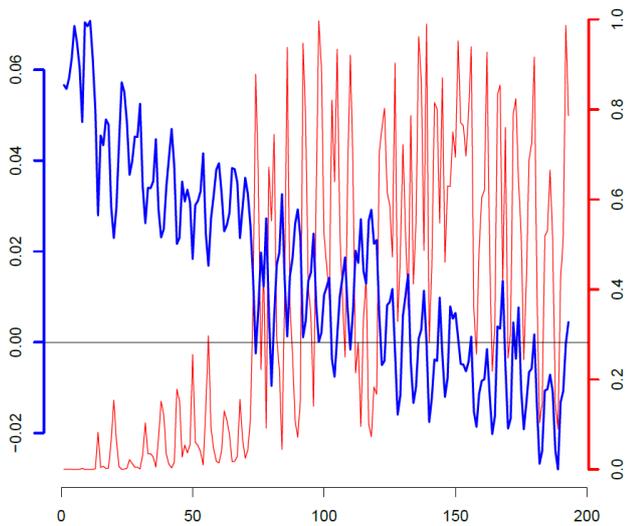

**Fig. 3**. Partial correlation plot between OMNI proton flux and AIM background on range Oct.10,2014-Apr.18,2015. AIM sources considered here have magnitude between 15 and 16. The left ordinate is the correlation value, while the right ordinate is the significance level (p-value). The abscissa value is the temporal shift, in hours. As you can expect, for time shifts of a few hours the partial correlation is significantly not zero, while increasing the time shift the correlation gets weaker.

Eventually time series can be pre-processed, through advanced moving averages or dimensionality reduction techniques like Principal Component Analysis (PCA), Independent Component Analysis (ICA), and Dynamic Factor Analysis (DFA); or SAX representation ([9]) that may highlight a common behaviour of processed series.

## 4. RESULTS

First processing has been a stress test, done on a significant time range of six months from October 10th, 2014 to April 18th, 2015. The time shift interval considered is one hour and two hundred shifts have been applied to data. The variables considered in this test are AIM background and flux along with OMNI proton flux, flow pressure, and plasma tempera-

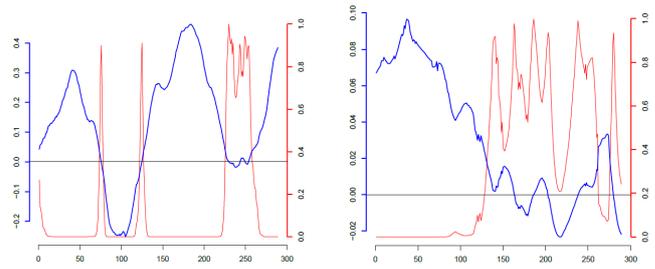

**Fig. 4**. Partial correlation between OMNI variables and AIM variables on Dec.12-25,2014. OMNI proton density vs. AIM (mag<13) flux; OMNI flow pressure vs. AIM (15<mag<16). background

ture and speed. The partial correlation results are shown in Fig. 3.

As Fig. 3 shows, the significant correlation is present inside the first hours of shifting; so there is effectively some correlation between Gaia data and interplanetary environment measurements. In order to refine this analysis, the following results use a shorter time shift interval (5 minutes) than above. Moreover, they will be restricted on days when some peaks of particles fluxes are present, to highlight correlation vs. delay behaviour. These days have been found making a scatterplot of the proton flux variable on the entire OMNI range considered (October 1, 2014 - May 31, 2015).

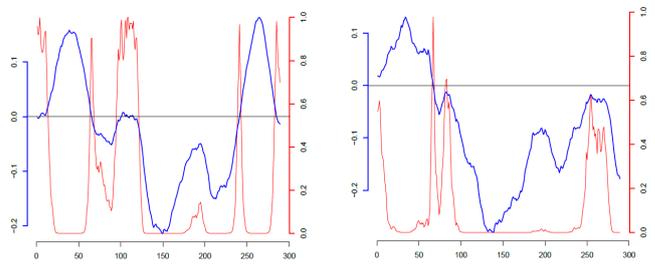

**Fig. 5**. Partial correlation between OMNI variables and AIM (mag<13) variables on May 11-15,2015. OMNI proton flux vs. AIM flux; OMNI proton flux vs. AIM background.

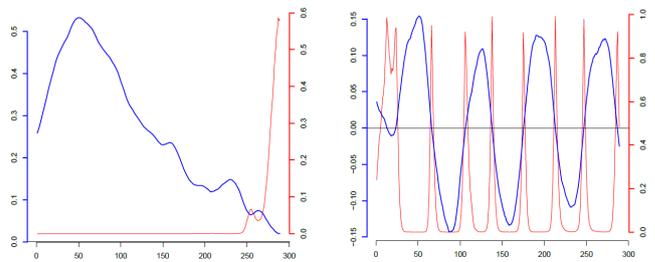

**Fig. 6**. Partial correlation between OMNI variables and AIM (15<mag<16) variables on May 11-15,2015. OMNI proton flux vs. AIM flux; OMNI proton flux vs. AIM background.

Considering days between December 12 and 25, 2014, when a bigger proton flux has been detected, some common patterns have been seen. In fact between 40 and 50 5-minutes delays a significant peak of correlation is found, as Fig. 4 shows. This happens for both the AIM magnitude ranges specified in Sec. 3.1.

Another interesting period for our analysis is the period between May 11 and 15, 2015. The plots about this period are in Fig. 5, as much as regards AIM data related to magnitude below 13, and Fig. 6 for magnitude between 15 and 16.

These plots highlight a correlation peak at about 45 iterations. This could mean that an higher particles activity shows its effect on Gaia almost four hours later, which could lead to a speed of 200 km/s from L1 to L2; this is compatible at least with the speed of MHD waves [10], [11]. However, further analyses are to be done with some major solar events happened after the time range we considered. Some refinements in the algorithms are also in progress to remove the periodicity from the plots, for example. This periodicity (of 6 hours, as you can see from plots) comes from the time it takes for Gaia to complete a rotation around its axis.

## 5. FURTHER WORK

Next steps of TECSEL2 project are both infrastructural and scientific. On one hand there is the optimization of the platform to perform data movements and analyses more efficiently, along with an improvement of interface usability.

On the other hand other time intervals containing bigger solar events are to be checked. Moreover AIM data on other CCD rows, and even other Gaia datasets may be used.

This processing can use the same algorithms introduced in Sec. 3.3, eventually with some improvements (e.g., removing time series periodicity) but also convolutive methods ([12]) or methods from topological analysis in order to study and infer the data structure from low dimensional representations.

## 6. CONCLUSIONS

TECSEL2 is an innovative project for several reasons. It is one of the first studies devoted to the monitoring and characterization of the behaviour of CCD detectors located in the L2 environment and it is therefore a key study for future space missions equipped with CCDs array similar to the one used on Gaia. As a service, TECSEL2 system is a powerful tool for efficient analysis of large and generic time series data, built with big data technologies, the state of art for the treatment of huge amount of data like those coming from the new generation of space and on-ground telescope.

Our results show a correlation between particles fluxes detected at L1 and charge flux and background detected on Gaia CCDs at L2 and processed by AIM. This correlation reaches its maximum with barely 4 hours of delay. Further analyses are in progress to investigate this effect.